\documentclass[12pt, draftclsnofoot, onecolumn]{IEEEtran}
\IEEEoverridecommandlockouts

\usepackage{amsmath,graphicx}
\usepackage{amssymb}
\usepackage{subfig}
\usepackage{graphicx}
\usepackage{graphics}
\usepackage{epstopdf}
\usepackage{amsmath}
\usepackage{multirow}
\usepackage{bm}
\usepackage{enumerate}
\usepackage{mathdots}
\usepackage{todonotes}
\usepackage{algorithmicx}
\usepackage{algpseudocode}
\usepackage{algorithm}
\usepackage{verbatim}
\usepackage{balance}
\usepackage{cite}
\usepackage{url}

\usepackage{color}
\definecolor{purple(x11)}{rgb}{0.63, 0.36, 0.94}
\definecolor{cadmiumgreen}{rgb}{0.0, 0.42, 0.24}

\newcommand{\var}{\mathop{\mathrm{var}}}
\newcommand{\Real}{\mathop{\mathrm{Re}}}
\newcommand{\Imag}{\mathop{\mathrm{Im}}}
\newcommand{\trace}{\mathop{\mathrm{trace}}}
\newcommand{\diag}{\mathop{\mathrm{diag}}}
\newcommand{\SNR}{\mathop{\mathrm{SNR}}}

\usepackage{pifont}
\newtheorem{lemma}{Lemma}

\newcommand{\Nr}{N_{\mathrm{r}}}
\newcommand{\Nt}{N_{\mathrm{t}}}
\newcommand{\Ns}{N_{\mathrm{s}}}
\newcommand{\Lt}{L_{\mathrm{t}}}
\newcommand{\Lr}{L_{\mathrm{r}}}

\newcommand{\Gr}{G_{\mathrm{r}}}
\newcommand{\Gt}{G_{\mathrm{t}}}

\newcommand{\ar}{{\mathbf{a}}_{\mathrm{R}}}
\newcommand{\at}{{\mathbf{a}}_{\mathrm{T}}}

\newcommand{\AR}{{\mathbf{A}}_{\mathrm{R}}}
\newcommand{\AT}{{\mathbf{A}}_{\mathrm{T}}}

\newcommand{\cH}{\mathbf{H}}
\newcommand{\FRF}{{\mathbf{F}}_{\mathrm{RF}}}
\newcommand{\FBB}{{\mathbf{F}}_{\mathrm{BB}}}
\newcommand{\WRF}{{\mathbf{W}}_{\mathrm{RF}}}
\newcommand{\WBB}{{\mathbf{W}}_{\mathrm{BB}}}

\newcommand{\be}{\begin{eqnarray}}
\newcommand{\ee}{\end{eqnarray}}

\def\ba{{\mathbf{a}}}

\def\bee{{\mathbf{e}}}

\def\bq{{\mathbf{q}}}
\def\br{{\mathbf{r}}}
\def\bs{{\mathbf{s}}}
\def\bt{{\mathbf{t}}}

\def\bv{{\mathbf{v}}}

\def\bx{{\mathbf{x}}}

\def\b0{{\mathbf{0}}}

\def\bA{{\mathbf{A}}}
\def\bB{{\mathbf{B}}}
\def\bC{{\mathbf{C}}}
\def\bD{{\mathbf{D}}}

\def\bF{{\mathbf{F}}}
\def\bG{{\mathbf{G}}}
\def\bH{{\mathbf{H}}}
\def\bI{{\mathbf{I}}}

\def\bM{{\mathbf{M}}}

\def\bP{{\mathbf{P}}}

\def\bW{{\mathbf{W}}}

\begin{document}

\title{Joint Data-Aided Carrier Frequency Offset, Phase Offset, Amplitude and SNR Estimation for Millimeter-Wave MIMO Systems}

\author{Javier Rodr\'{i}guez-Fern\'{a}ndez$^{\dag}$, Nuria Gonz\'{a}lez-Prelcic$^{\dag}$, and Robert W. Heath Jr.$^{\ddag}$
\\
$^\dag$ Universidade de Vigo, Email: $\{$jrodriguez,nuria$\}$@gts.uvigo.es \\
$^\ddag$ The University of Texas at Austin, Email: $\{$kiranv,rheath$\}$@utexas.edu}

\maketitle
\begin{abstract}
This work is devoted to solve the problem of estimating the carrier frequency offset, phase offset, amplitude, and SNR between two mmWave transceivers.

 The Cram\'{e}r-Rao Lower Bound (CRLB) for the different parameters is provided first, as well as the condition for the CRLB to exist, known as Regularity Condition. Thereafter, the problem of finding suitable estimators for the parameters is adressed, for which the proposed solution is the Maximum Likelihood estimator (ML).
\end{abstract}
 
 \textbf{Notation}: here we describe the notation to be used throughout this document: bold uppercase $\bA$ is used to denote matrices, bold lowercase $\ba$ denotes a column vector and non-bold lowercase $a$ denotes a scalar value. We use ${\cal A}$ to denote a set. Further, $\|\bA\|_F$ is the Frobenius norm and $\bA^*$, $\bA^\text{C}$, $\bA^T$ and $\bA^\dag$ denote the conjugate transpose, conjugate, transpose and Moore-Penrose pseudoinverse of a matrix $\bA$, respectively. The $(i,j)$-th entry of a matrix $\bA$ is denoted using $[\bA]_{i,j}$. Similarly, the $i$-th entry of a column vector $\ba$ is denoted as $[\ba]_i$. The identity matrix of order $N$ is denoted as $\bI_N$. If $\bA$ and $\bB$ are two matrices, $\bA \circ \bB$ is the Khatri-Rao product of $\bA$ and $\bB$ and $\bA \otimes \bB$ is their Kronecker product. We use ${\cal N}(\bm{\mu},\bC)$ to denote a circularly symmetric complex Gaussian random vector with mean $\bm{\mu}$ and covariance matrix $\bC$. We use $\mathbb{E}$ to denote expectation. Discrete-time signals are represented as $\bx[n]$.
 
\section{System Model}
\begin{figure*}[t!]
\centering
\includegraphics[width=0.8\textwidth]{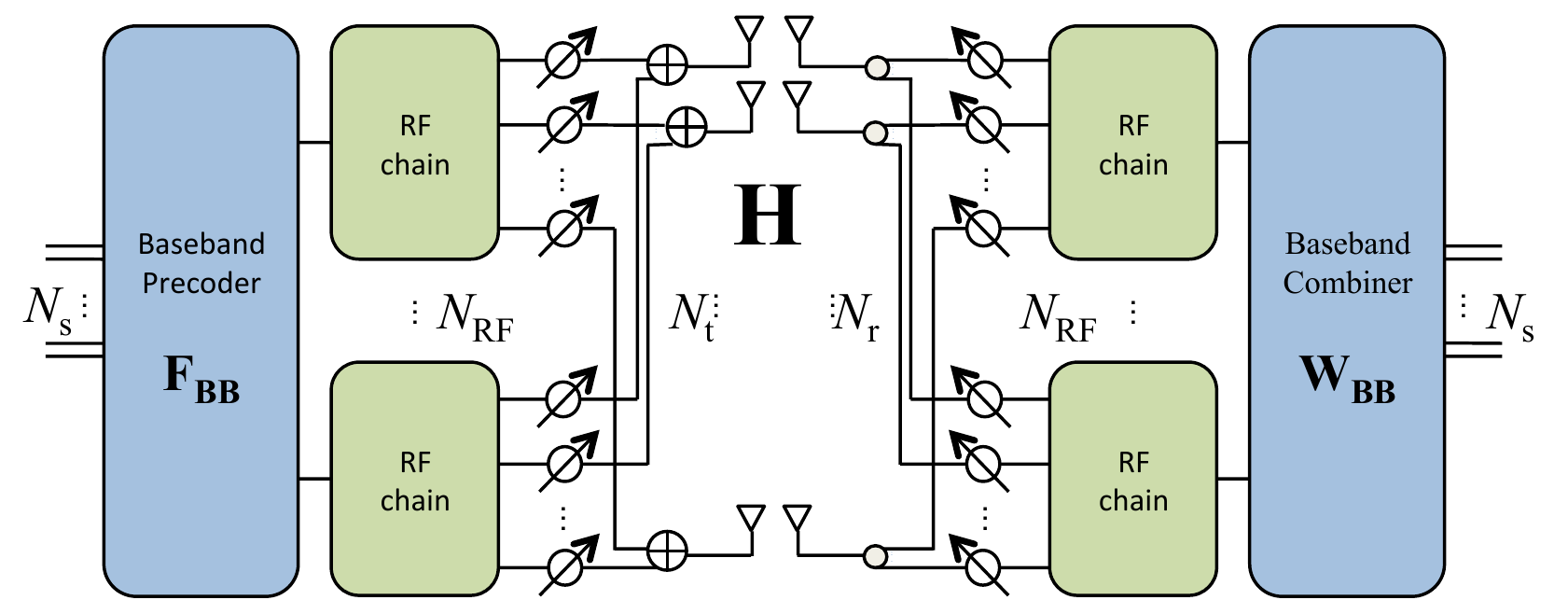} 
\caption{Illustration of the structure of a hybrid MIMO architecture, which include analog and digital precoders and combiners.}     
\label{fig:hybrid_architecture}        
\end{figure*}

We consider a mmWave MIMO link to send $\Ns$ data streams using a transmitter with $\Nt$ antennas and a receiver with $\Nr$ antennas. 
Both the transmitter and the receiver use a fully-connected hybrid MIMO architecture as shown in Fig. \ref{fig:hybrid_architecture}, with $\Lr$ and $\Lt$ RF chains. A hybrid precoder is used, with $\mathbf{F}= \FRF\FBB \in {\mathbb{C}}^{\Nt\times\Ns}$, where $\FRF \in \mathbb{C}^{\Nt \times \Lt}$ is the analog precoder and $\FBB \in \mathbb{C}^{\Lt \times \Ns}$ the digital one. The RF precoder and combiner are implemented using a fully connected network of phase shifters, as described in \cite{RiaRusPreAlkHea:Hybrid-MIMO-Architectures:16}. 

The MIMO channel between the transmitter and the receiver is modeled as a $\Nr \times \Nt$ matrix denoted as $\bH$, which is assumed to be a sum of the contributions of $C$ spatial clusters, each contributing with $R_c$ rays. This matrix is given by \cite{RiaRusPreAlkHea:Hybrid-MIMO-Architectures:16}
\be
\hspace*{-3.5mm}\cH &=\hspace*{-1mm}& \sqrt{\frac{\Nr\Nt}{\rho_\text{L} \sum_{c=1}^{C}{R_c}}}\sum_{c = 1}^{C}{\sum_{r=1}^{R_c}{\alpha_{c,r}\ar(\phi_{c,r})\at^*(\theta_{c,r})}}, \label{eqn:channel_model}
\ee
where $\rho_\text{L}$ denotes the path loss, $C$ is the number of scattering clusters, $R_c$ is the number of rays for $c$-th cluster, $\alpha_{c,r} \in {\mathbb{C}}$ is the complex gain of the $r$-th ray within $c$-th cluster, $\phi_{c,r} \in [0, 2\pi)$ and $\theta_{c,r} \in [0, 2\pi)$ are the angles of arrival and departure (AoA/AoD), respectively of the $r$-th ray within $c$-th cluster, and $\ar(\phi_{c,r}) \in {\mathbb{C}}^{\Nr\times1}$ and $\at(\theta_{c,r}) \in {\mathbb{C}}^{\Nt\times1}$ are the array steering vectors for the receive and transmit antennas respectively.  This matrix can be written in a more compact way as
\be
\cH &=& \AR \bG \AT^*, \label{eqn:channel_compact}
\ee where $\bG \in {\mathbb{C}}^{\sum_{c=1}^{C}{R_c}\times \sum_{c=1}^{C}{R_c}}$ is diagonal with non-zero complex entries, %$\sqrt{\frac{\Nt\Nt}{L}}\alpha_{\ell}\prc(dT_s-\tau_{\ell})$, 
and $\AR \in {\mathbb{C}}^{\Nr\times \sum_{c=1}^{C}{R_c}}$ and $\AT \in {\mathbb{C}}^{\Nt\times \sum_{c=1}^{C}{R_c}}$ contain the receive and transmit array steering vectors $\ar(\phi_{c,r})$ and $\at(\theta_{c,r})$, respectively. $\cH_d$ can be approximated using the extended virtual channel representation \cite{mmWavetutorial} as
\begin{equation}
			\cH \approx
				 \tilde{\bA}_\text{R} \bG^\text{v} \tilde{\bA}_\text{T}^*
\label{eq:virtual_ch_model}
\end{equation}
where  $\bG^\text{v} \in \mathbb{C}^{\Gr \times \Gt}$ is a sparse matrix which contains the path gains of the quantized spatial frequencies in the non zero elements. The dictionary matrices 
$\tilde{\bA}_\text{T}$ and $\tilde{\bA}_\text{R}$  contain the transmitter and receiver array response vectors evaluated on grids of sizes $\Gt$ and $\Gr$. 
Assuming that the receiver applies a hybrid combiner ${\mathbf{W}}={\WRF\WBB} \in {\mathbb{C}}^{\Nr\times\Lr}$, with $\WRF \in \mathbb{C}^{\Nr \times \Lr}$ the analog combiner, and $\WBB \in \mathbb{C}^{\Lr \times \Ns}$ the baseband combiner, the received signal at discrete time instant $n$ can be written as
\be
\br[n]  =  \bW^*\bH\bF\bs[n] e^{j2\pi \Delta f n} + \bv[n], 
\label{equation:rx_signal}
\ee
for $n = 0,\ldots,N-1$.The signal $\bs[n] \in \mathbb{C}^{\Lt \times 1}$, $0\leq n \leq N-1$ is a training sequence known to the receiver, $\Delta f$ is the unknown carrier frequency offset (CFO), and $\bv[n] \sim \mathcal{N}\left(0,\sigma^2 \bW^*\bW\right)$  is the circularly symmetric complex Gaussian distributed additive noise vector.

We consider that the training sequence can be expressed as $\bs[n] = \bq t[n]$, with $\bq \in \mathbb{C}^{\Lt \times 1}$ a spatial filter consisting of normalized QPSK symbols. Let us define $\bm \alpha \triangleq \bD_\text{w}^{-*}\bW^{*}\bH \bF \bq$, $\bm \alpha = \left[\begin{array}{cccc} \alpha_1 e^{j\beta_1} & \alpha_2 e^{j\beta_2} & \ldots & \alpha_{L_\text{r}} e^{j\beta_{L_\text{r}}} \\ \end{array}\right]^T$ and $\bm \Omega_n = e^{j2\pi \Delta f n}\bI_{\Lr}$, where $\bD_\text{w} \in \mathbb{C}^{\Lr \times \Lr}$ is the Cholesky factor of $\bC_\text{w} = \bW^{*}\bW$, i.e., $\bC_\text{w} = \bD_\text{w}^{*}\bD_\text{w}$. Then, if we stack the $N$ samples of the received signal in \eqref{equation:rx_signal} we obtain the signal model
\begin{equation}
	\overbrace{\left[\begin{array}{c} \br[0] \\
	\br[1] \\
	\vdots \\
	\br[N-1] \\ \end{array}\right]}^{\br} = \overbrace{\left[\begin{array}{cccc} \bm \Omega_0 \bm \alpha & & & \\
	& \bm \Omega_1 \bm \alpha & & \\
	& & \ddots & \\
	& & & \bm \Omega_{N-1}\bm \alpha \\ \end{array}\right]}^{\bM(\bm \alpha, \Delta f)} \overbrace{\left[\begin{array}{c} t[0] \\ t[1] \\ \vdots \\ t[N-1] \\ \end{array}\right]}^{\bt} + \overbrace{\left[\begin{array}{c} \bv[0] \\ \bv[1] \\ \vdots \\ \bv[N-1] \\ \end{array}\right]}^{\bv},
	\label{equation:stacked_signal_model}
\end{equation}
such that $\br \sim {\cal N}(\bM(\bm \alpha,\Delta f) \bt, \sigma^2 \bI_{M\Lr})$. Then, the vector of parameters to be estimated is $\bm \xi = \left[\begin{array}{cccccccc} \alpha_1 & \ldots & \alpha_{\Lr} & \beta_1 & \ldots & \beta_{\Lr} & \Delta f & \sigma^2 \\ \end{array}\right]^T$.

We define the Signal to Noise Ratio (SNR) for the $i$-th signal $[\br[n]]_i$, $i = 1, \ldots, \Lr$ as
\begin{equation}
\gamma_i = \frac{\alpha_i^2}{\sigma^2}.
\end{equation}
Furthermore, the average Signal to Noise Ratio at digital level can be defined as well. Let us define $\bm \Lambda\in \mathbb{C}^{\Lr\times \Lr}$ as $\bm \Lambda = \diag\{\alpha_1e^{j\beta_1},\alpha_2e^{j\beta_2},\ldots, \alpha_{\Lr}e^{j\beta_{\Lr}}\}$. Therefore, if the SNR at baseband level is denoted as $\SNR$, it can be written as
\begin{equation}
\SNR = \frac{1}{\Lr}\sum_{i=1}^{\Lr}{\gamma_i} = \frac{\trace\{\bm \Lambda \bm \Lambda^*\}}{\Lr\sigma^2},
\end{equation}
which is just the average of the post-combining $\gamma_i$ at each RF chain.
 
 \section{Regularity Condition}
 
 For the CRLB to exist, the regularity condition must be fulfilled by the probability density function (pdf) of the data. This condition states \cite{Kay:Fundamentals-of-Statistical-Signal:93}
 \begin{equation}
 \mathbb{E}{\bigg{\{}}\frac{\partial \ln{p(\br; \bm \xi)}}{\partial \xi_i}{\bigg{\}}} = 0,\qquad \text{for all }\xi_i \in \bm \xi.
 \end{equation}
 The pdf of the vector $\br$ is written as
 \begin{equation}
 p(\br; \bm \xi) = \frac{1}{\pi^{N\Lr} \det{\big{(}}\sigma^2 \bI_{N\Lr}{\big{)}}}e^{-\frac{1}{\sigma^2}(\br - \bM(\bm \alpha,\Delta f) \bt )^*(\br - \bM(\bm \alpha,\Delta f) \bt)},
 \end{equation}
 and the log-likelihood function (LLF) as
 
 \begin{equation}
 \begin{split}
 \ln{p(\br; \bm \xi)} &= -N\Lr\ln{\pi \sigma^2} - \frac{1}{\sigma^2}{\bigg{(}}\br^* \br - 2\Real\{\bt^{*} \bM^*(\bm \alpha,\Delta f) \br\}+ \\ &+ \bt^{*} \bM^*(\bm \alpha,\Delta f) \bM(\bm \alpha,\Delta f) \bt{\bigg{)}}.
 \end{split}
 \end{equation}
 
 Before computing the gradient of the LLF, it is important to note that
 
 \begin{equation}
 \begin{split}
 \bt^{*} \bM(\bm \alpha,\Delta f) \bM(\bm \alpha,\Delta f) \bt &= \sum_{n=0}^{N-1}{s^\text{C}[n] \bm \alpha^* \bm \Omega_n^* \bm \Omega_n \bm \alpha s[n]} \\ &= \sum_{n=0}^{N-1}{ s^\text{C}[n] \trace\{\bP\}  s[n]} \\ &= N\trace\{\bP\},
 \end{split}
 \end{equation}
 where $\bP = \bm \Lambda \bm \Lambda^*$.
 
 Therefore, for $\alpha_i$,
 
 \begin{equation}
 \begin{split}
 \frac{\partial \ln{p(\br; \bm \xi)}}{\partial \alpha_i} &= -\frac{1}{\sigma^2}{\bigg{(}}-2\Real\left\{\bt^{*} \frac{\partial \bM^*(\bm \alpha,\Delta f)}{\partial \alpha_i} \br\right\} +N \Lr \frac{\partial{\trace\{\bP\}}}{\partial \alpha_i}{\bigg{)}} \\ &= -\frac{1}{\sigma^2}{\bigg{(}}-2\sum_{n=0}^{N-1}{\Real\left\{s^\text{C}[n] e^{-j\beta_i} \bee_i^T \bm \Omega_n^* \br[n]\right\}} +2\alpha_i N {\bigg{)}} \\
 &= -\frac{1}{\sigma^2}{\bigg{(}}-2\sum_{n=0}^{N-1}{\Real\left\{s^\text{C}[n] e^{-j\beta_i}\bee_i^T \bm \Omega_n^* \bm \Omega_n \bm \alpha  s[n]\right\}}+2\alpha_i N {\bigg{)}} \\
 &=\mathbb{E}{\bigg{\{}}\frac{\partial \ln{p(\br; \bm \xi)}}{\partial \alpha_i}{\bigg{\}}} = -\frac{1}{\sigma^2}{\bigg{(}}-2\sum_{n=0}^{N-1}{\Real\{\alpha_i\}} + 2\alpha_i N{\bigg{)}} = 0.
 \end{split}
 \end{equation}

 Now, for $\sigma^2$,
 
 \begin{equation}
 \begin{split}
 \frac{\partial \ln{p(\br; \bm \xi)}}{\partial \sigma^2} &= -\frac{N\Lr}{\sigma^2} + \frac{1}{\sigma^4}\|\br - \bM(\bm \alpha,\Delta f) \bt \|_2^2 \\
	&= -\frac{N\Lr}{\sigma^2} + \frac{1}{\sigma^4}\sum_{n=0}^{N-1}{\mathbb{E}{\big{\{}}\|\bm w_n[n]\|^2{\big{\}}}} \\
 &= - \frac{N\Lr}{\sigma^2} + \frac{1}{\sigma^4}\sum_{n=0}^{N-1}{\sum_{i=1}^{\Lr}{{\big{(}}\overbrace{\mathbb{E}\{w_{I,i}[n]^2\} + \mathbb{E}\{w_{Q,i}[n]^2}^{\sigma^2}{\big{)}}\}}} = 0.
 \end{split}
 \end{equation}
 
 For $\Delta f$,
 
 \begin{equation}
 \begin{split}
 \frac{\partial \ln{p(\br; \bm \xi)}}{\partial \Delta f} &= -\frac{1}{\sigma^2}{\bigg{(}}-2\Real\left\{\bt^{*} \frac{\partial \bM^*(\bm \alpha,\Delta f)}{\partial \Delta f}\br \right\}{\bigg{)}} \\
 &= -\frac{1}{\sigma^2}{\bigg{(}}-2\sum_{n=0}^{N-1}{\Real\left\{s^\text{C}[n]\bm \alpha^* (-j2\pi n) \bm \Omega_n^* \bm \Omega_n \bm \alpha s[n]\right\}}{\bigg{)}} \\
 &= -\frac{1}{\sigma^2}{\bigg{(}}-2\sum_{n=0}^{N-1}{\Real\left\{\trace\{\bP\} (-j2\pi n) \right\}}{\bigg{)}} = 0
	\end{split} 
 \end{equation}

 Finally, for $\beta_i$,
 
 \begin{equation}
 \begin{split}
 \frac{\partial \ln{p(\br; \bm \xi)}}{\partial \beta_i} &= -\frac{1}{\sigma^2}{\bigg{(}}-2\Real\left\{\bt^{*} \frac{\partial \bM^*(\bm \alpha,\Delta f)}{\partial \beta_i} \br\right\}{\bigg{)}} \\
  &= -\frac{1}{\sigma^2}{\bigg{(}}-2\sum_{n=0}^{N-1}{\Real\left\{s^\text{C}[n] \alpha_i(-j) e^{-j\beta_i} \bee_i^T \bm \Omega_n^* \bm \Omega_n \bm \alpha s[n] \right\}}{\bigg{)}} \\
  &= -\frac{1}{\sigma^2}{\bigg{(}}-2\sum_{n=0}^{N-1}{\Real\{\alpha_i^2(-j) \}}{\bigg{)}} = 0,
  \end{split}
 \end{equation}
 
 since the term inside brackets is also purely imaginary.
 
 Thus, since the regularity condition holds, the CRLB is to be found in the next section.
 
 \section{Cram\'{e}r-Rao Lower Bound}
 
 Since the model for the received signal is Gaussian, the Slepian-Bangs formula can be used to find the elements in the Fisher Information Matrix. This formula is given by \cite{Kay:Fundamentals-of-Statistical-Signal:93}
 
 \begin{equation}
 \bF_{i,j}(\bm \xi) = 2\Real{\bigg{\{}}\frac{\partial \bm \mu^*(\bm \xi)}{\partial \xi_i}\bm C^{-1}(\bm \xi)\frac{\partial \bm \mu(\bm \xi)}{\partial \xi_j}{\bigg{\}}} + \trace{\bigg{\{}}\bm C^{-1}(\bm \xi)\frac{\partial \bm C(\bm \xi)}{\partial \xi_i}\bm C^{-1}(\bm \xi)\frac{\partial \bm C(\bm \xi)}{\partial \xi_j}{\bigg{\}}}.
 \end{equation}
 
Thus, the elements in the FIM can be found as
 
 \begin{equation}
 \bF_{\alpha_i,\alpha_i}(\bm \xi) = \frac{2}{\sigma^2}\Real{\bigg{\{}}\bt^{*} \frac{\partial \bM^*(\bm \alpha,\Delta f)}{\partial \alpha_i}\frac{\partial \bM(\bm \alpha,\Delta f)}{\partial \alpha_i}\bt{\bigg{\}}}.
 \end{equation}
 
The partial derivative of $\bM(\bm \alpha,\Delta f)$ is given by
 
 \begin{equation}
 \frac{\partial \bM^*(\bm \alpha,\Delta f)}{\partial \alpha_i}\frac{\partial \bM(\bm \alpha,\Delta f)}{\partial \alpha_i} =\left[\begin{array}{cccc}  \bee_i^T   \bee_i & & & \\
 & \bee_i^T  \bee_i & & \\
 & & \ddots & \\
 & & & \bee_i^T  \bee_i \end{array}\right] = \bI_N,
 \end{equation}
 
 being $\bee_i \in \mathbb{R}^{\Lr}$ a vector of zeros with a single one in the $i$-th position. Therefore, 
 
 \begin{equation}
 \bF_{\alpha_i,\alpha_i}(\bm \xi) = \frac{2}{\sigma^2}\sum_{n=0}^{N-1}{\Real\{|s[n]|^2\}} = \frac{2N}{\sigma^2}.
 \end{equation}
 
 For the phase offset parameter,
 
 \begin{equation}
 \bF_{\beta_i,\beta_i}(\bm \xi) = \frac{2}{\sigma^2} \Real{\bigg{\{}}\bt^{*} \frac{\partial \bM^*(\bm \alpha,\Delta f)}{\partial \beta_i} \frac{\partial \bM(\bm \alpha,\Delta f)}{\partial \beta_j}\bt{\bigg{\}}},
 \end{equation}
 
 where
 
 \begin{equation}
 \frac{\partial \bM^*(\bm \alpha,\Delta f)}{\partial \beta_i} \frac{\partial \bM(\bm \alpha,\Delta f)}{\partial \beta_i} = \left[\begin{array}{cccc}  \alpha_i^2\bee_i^T \bee_i & & & \\
 & \alpha_i^2\bee_i^T  \bee_i & & \\
 & & \ddots & \\
 & & & \alpha_i^2 \bee_i^T \bee_i \end{array}\right] = \alpha_i\bI_N.
 \end{equation}
 
 Then, the Fisher Information for $\beta_i$ yields
 
 \begin{equation}
 \bF_{\beta_i,\beta_i}(\bm \xi) = \frac{2}{\sigma^2} \sum_{n=0}^{N-1}{\Real\{\alpha_i^2 |s[n]|^2\}} = \frac{2N\alpha_i^2}{\sigma^2}.
 \end{equation}
 
 For the carrier frequency offset parameter,
 
 \begin{equation}
 \bF_{\Delta f, \Delta f}(\bm \xi) = \frac{2}{\sigma^2}\Real{\bigg{\{}}\bt^{*} \frac{\partial \bM^*(\bm \alpha,\Delta f)}{\partial \Delta f}\frac{\partial \bM(\bm \alpha,\Delta f)}{\partial \Delta f}\bt{\bigg{\}}},
 \end{equation}
 
 where
 
 \begin{equation}
 \frac{\partial \bM^*(\bm \alpha,\Delta f)}{\partial \Delta f}\frac{\partial \bM(\bm \alpha,\Delta f)}{\partial \Delta f} = \trace\{\bP\}\left[\begin{array}{cccc}  (2\pi 0)^2  & & & \\
 & (2\pi 1)^2  & & \\
 & & \ddots & \\
 & & & (2\pi (N-1))^2   \end{array}\right].
 \end{equation}
 
 Therefore,
 
 \begin{equation}
 \bF_{\Delta f, \Delta f}(\bm \xi) = \frac{2}{\sigma^2}\trace\{\bP\}\sum_{n=0}^{N-1}{(2\pi n)^2}.
 \end{equation}
 
 For the noise variance,
 
 \begin{equation}
 \bF_{\sigma^2,\sigma^2}(\bm \xi) = \frac{1}{\sigma^4}\trace\{\bI_{N\Lr}\} = \frac{N\Lr}{\sigma^4}.
 \end{equation}
 
 For the non-diagonal elements in the FIM, it can be checked that all of them are zero-valued except for
 
 \begin{equation}
 \bF_{\Delta f, \beta_i}(\bm \xi) = \frac{2}{\sigma^2}\Real{\bigg{\{}}\bt^{*} \frac{\partial \bM^*(\bm \alpha,\Delta f)}{\partial \Delta f}\frac{\partial \bM(\bm \alpha,\Delta f)}{\partial \beta_i}\bt{\bigg{\}}},
 \end{equation}
 
 where 
 
 \begin{equation}
	 \frac{\partial \bM^*(\bm \alpha,\Delta f)}{\partial \Delta f}\frac{\partial \bM(\bm \alpha,\Delta f)}{\partial \beta_i} = \alpha_i e^{j\beta_i}\left[\begin{array}{cccc}  (2\pi 0)  \bm \alpha^* \bee_i & & & \\
	 & (2\pi 1) \bm \alpha^* \bee_i & & \\
	 & & \ddots & \\
	 & & & (2\pi (N-1))  \bm \alpha^* \bee_i \end{array}\right].
 \end{equation}
 
 Therefore,
 
 \begin{equation}
	 \bF_{\Delta f, \beta_i}(\bm \xi) = \frac{2}{\sigma^2}\sum_{n=0}^{N-1}{\Real\{(2\pi n) \alpha_i^2\}} = \frac{2\alpha_i^2}{\sigma^2}\sum_{n=0}^{N-1}{(2\pi n)}.
 \end{equation}

 Then, the FIM is found to be
 
 \begin{equation}
 \bF (\bm \xi) = \left[\begin{array}{ll} \bF_1(\bm \xi) & \bm 0 \\
 \bm 0 & \bF_2(\bm \xi) \end{array}\right],
 \end{equation}
 
 with 
 
 \begin{equation}
 \bF_1(\bm \xi) = \diag\left\{\frac{2N}{\sigma^2}\bm 1_{\Lr}^T, \frac{\Lr N}{\sigma^4}\right\}
 \end{equation}
 
 \begin{equation}
 \bF_2(\bm \xi) = \left[\begin{array}{ll}
 \frac{2\trace\{\bP\}}{\sigma^2} \sum_{n=0}^{N-1}{(2\pi n)^2} & \frac{2}{\sigma^2}\sum_{n=0}^{N-1}{2\pi n}\bm 1_{\Lr}^T \bP \\
 \frac{2}{\sigma^2}\sum_{n=0}^{N-1}{2\pi n}\bP\bm 1_{\Lr} & \frac{2N}{\sigma^2} \bP \\
 \end{array}\right].
 \end{equation}

 Finally, upon inverting the block diagonal matrix $\bF(\bm \xi)$, the CRLB for the parameters is given by
 
 \begin{equation}
 \var\{\hat \alpha_i\} \geq \frac{\sigma^2}{2N}
 \end{equation}
 
 \begin{equation}
 \var\{\widehat{\sigma^2}\} \geq \frac{\sigma^4}{\Lr N}
 \end{equation}
 \footnotesize
 \begin{equation}
 \var\{\widehat{ \Delta f}\} \geq \frac{\sigma^2}{2}\cdot\frac{N^{\Lr}}{{{\big{(}}N(N-1)(2N-1)/6{\big{)}} \trace\{\bP\} N^{\Lr} - \sum_{k=1}^{\Lr}{\alpha_k^2 {\big{(}}N(N+1)^2/4{\big{)}} N^{\Lr}  }}}
 \end{equation}
 \footnotesize
 \begin{equation}
 \var\{\hat \beta_i\} \geq \frac{\sigma^2}{2} \cdot \frac{{\big{(}}(N-1)(2N-1)/6{\big{)}} \trace\{\bP\}N^{\Lr}\prod_{\substack{j=1\\j\neq i}}^{\Lr}{\alpha_j^2} - \sum_{\substack{k=1\\k \neq i}}^{\Lr}{\alpha_k^4{\big{(}}(N+1)^2/4{\big{)}} N^{\Lr}\prod_{\substack{s=1\\s\neq k,i}}^{\Lr}{\alpha_s^2}}}{{\big{(}}N(N-1)(2N-1)/6{\big{)}} \trace\{\bP\} N^{\Lr} - \sum_{k=1}^{\Lr}{\alpha_k^2 {\big{(}}N(N+1)^2/4{\big{)}} N^{\Lr}  }}
 \end{equation}
 \normalsize
 Finally,  the $\SNR$ in the $i$-th RF chain can be distinguished from the average of the individual $\SNR$s of the different signals involved in the problem. If the $\SNR$ of the $i$-th signal is denoted by $\gamma_i$, then
 
 \begin{equation}
 \gamma_i = \frac{\alpha_i^2}{\sigma^2}
 \end{equation}
 
 \begin{equation}
 \SNR = \frac{\trace\{\bP\}}{\Lr\sigma^2},
 \end{equation}
 
 for which, according to the \textit{Transformed Parameters Property}, the CRLB can be found to be
 
 \begin{equation}
 \var\{\hat \gamma_i\} \geq \frac{4\gamma_i}{N}+\frac{\gamma_i^2}{N}
 \end{equation}
 \begin{equation}
 \var\{\widehat{\SNR}\} \geq \frac{2\SNR}{\Lr N} + \frac{\SNR^2}{\Lr N}.
 \end{equation}
 
 The last formulas provide a clear insight on how the estimation of $\gamma_i$ actually works. For lower values of $\gamma_i$, the CRLB is dominated by the term growing linearly with the parameter, whereas for higher values the CRLB is dominated by the second term, making the estimation of the parameter much harder. The same happens for $\widehat{\SNR}$, although it actually depends on the different $\gamma_i$ for the different streams.
 
 Henceforth, our effort is focused on searching for suitable estimators of these parameters. Owing to the non-linear dependence between the data and the parameters, an efficient estimator cannot be found, in general. 
 The practical approach to follow is to seek the ML estimators for these parameters, for which the next subsection is devoted.
 
 \section{Maximum Likelihood Estimation}
 
 The problem of finding the ML Estimator for the parameters in $\bm \xi$ can be formalized as
 \begin{equation}
 \underset{i = 1,\ldots,\Lr}{\left\{\hat {\alpha}_{i,\text{ML}}, \hat \sigma_\text{ML}^2, \hat \Delta f_\text{ML}, \hat {\beta}_{i,\text{ML}}\right\}} = \arg\,\underset{\bm \xi}{\max}\, \ln{p(\br; \bm \xi)},
 \end{equation}
 which involves a joint maximization over $2(\Lr+1)$ scalar variables. This problem can be solved in four different steps by splitting the original problem into four interconnected optimization problems.
 \subsection{MLE for the Phase Offset}
 
 Recall that the LLF is given by
 \begin{equation}
 \ln{p(\br; \bm \xi)} = -N\Lr\ln{\pi\sigma^2} - \frac{1}{\sigma^2}{\bigg{(}}\br^*\br - 2\Real\{\bt^{*} \bM^*(\bm \alpha,\Delta f) \br\} + N\trace\{\bP\}{\bigg{)}},
 \end{equation}
 such that the term that depends on $\beta_i$ is the second variable inside brackets. Therefore, the value $\hat \beta_i$ that maximizes the function above is found from
 \begin{equation}
 \hat \beta_i = \arg\,\underset{\beta_i}{\max}\quad \Real\{\bt^{*} \bM^*(\bm \alpha,\Delta f) \br\}.
 \end{equation}
 The first derivative of the function to maximize yields
 \begin{equation}
 \frac{\partial \ln{p(\br; \bm \xi)}}{\partial \beta} = \sum_{n=0}^{N-1}{\Real\{s^\text{C}[n]\alpha_i (-j) e^{-j\beta_i} \bee_i^T\bm \Omega_n \br[n]\}}
 \end{equation}
 \begin{equation}
 \frac{\partial \ln{p(\br; \bm \xi)}}{\partial \beta} = \sum_{n=0}^{N-1}{\Imag\{s^\text{C}[n]\alpha_i e^{-j\beta_i} e^{-j2\pi \Delta f n} r_i[n]\}}
 \end{equation}
 Now the following result can be applied.
 
 \textit{\textbf{Lemma}}. Given two complex numbers, $z_1$, $z_2$, the imaginary part of their product is defined as
 \begin{equation*}
 \Imag\{z_1z_2\} = \Real\{z_1\}\Imag\{z_2\} + \Imag\{z_1\}\Real\{z_2\}.
 \end{equation*}
 Therefore, setting the previous derivative to zero allows us to obtain
 \begin{equation}
 \sum_{n=0}^{N-1}{\Imag\{s^\text{C}[n]r_i[n] e^{-j2\pi \Delta f n}\}} \Real\{\alpha_i e^{-j\beta_i}\} =  \sum_{n=0}^{N-1}{\Real\{s^\text{C}[n]r_i[n] e^{-j2\pi \Delta f n}\}} \Imag\{\alpha_i e^{-j\beta_i}\}
 \end{equation}
 \begin{equation}
 \hat \beta_i = \tan^{-1}{\bigg{\{}}\frac{\sum_{n=0}^{N-1}{\Imag\{s^\text{C}[n]r_i[n] e^{-j2\pi \Delta f n}\}}}{\sum_{n=0}^{N-1}{\Real\{s^\text{C}[n]r_i[n] e^{-j2\pi \Delta f n}\}}}{\bigg{\}}},
 \label{equation:phi_ML}
 \end{equation}
 which can be interpreted as a matched-filtering operation with  the training sequence $s[n]$ after the carrier frequency offset has been corrected. Notice that the ML estimator of $\beta_i$ requires the knowledge of the true value $\Delta f$. Since it is impossible to know the exact value of $\Delta f$, the ML estimator of $\Delta f$ is to be substituted in \eqref{equation:phi_ML} instead, such that the final estimator can be applied in a practical scenario.
 
 \subsection{MLE for the Amplitude}
 
 In this subsection we provide a closed-form expression for the amplitude parameters, which have already been denoted as $\alpha_i$. By inspection of the LLF
 \begin{equation}
 \ln{p(\br; \bm \xi)} = -N\Lr\ln{\pi\sigma^2} - \frac{1}{\sigma^2}{\bigg{(}}\br^*\br - 2\Real\{\bt^{*} \bM^*(\bm \alpha,\Delta f) \br\} + N\trace\{\bP\}{\bigg{)}},
 \label{equation:LLF}
 \end{equation}
 it can be seen that the terms that depend on this parameter are the two last within brackets. Therefore, the problem of finding $\alpha_i$ is formalized as
 \begin{equation}
 \hat \alpha_i = \arg\,\underset{\alpha_i}{\max}\quad 2\Real\{\bt^{*} \bM^*(\bm \alpha,\Delta f) \br\} - N\trace\{\bP\}.
 \end{equation}
 Taking the first derivative of the objective function yields
 \begin{equation}
 \frac{\partial f(\alpha_i)}{\partial \alpha_i} = 2\sum_{n=0}^{N-1}{\Real\{r_i[n] s^\text{C}[n]e^{-j\beta_i} e^{-j2\pi \Delta f n}\}} - 2\alpha_i N,
 \end{equation}
 such that the value of $\alpha_i$ that maximizes the function above is 
 \begin{equation}
 \hat \alpha_i = \frac{1}{N}\sum_{n=0}^{N-1}{\Real\{r_i[n]s^\text{C}[n]e^{-j\beta_i} e^{-j2\pi \Delta f n}\}}.
 \label{equation:Amplitude_Estimator}
 \end{equation}
 From \eqref{equation:Amplitude_Estimator}, it is clear that the statistic $\hat \alpha_i$ depends on $\Delta f$ and $\beta_i$, similarly as with the estimator of $\beta_i$. Therefore, the estimators of these parameters are to be substituted in \eqref{equation:Amplitude_Estimator} to find the ML estimator. 
 
 \subsection{MLE for the Carrier Frequency Offset}
 
	In this subsection we will find the ML estimator of the carrier frequency offset. In \eqref{equation:LLF}, the only term that depends on $\Delta f$ is the second one. Therefore, the problem of finding the ML estimator for this parameter can be stated as
 \begin{equation}
 \widehat {\Delta f} = \arg\,\underset{\Delta f}{\max}\,\Real\{\bt^{*}\bM^*(\bm \alpha,\Delta f)\br\}.
 \label{equation:opt_CFO}
 \end{equation}
 We can express the objective function in \eqref{equation:opt_CFO} as
 \begin{equation}
 f(\Delta f) = \Real{\bigg{\{}}\sum_{n=0}^{N-1}{\sum_{i=1}^{\Lr}{\alpha_i r_i[n] s^\text{C}[n] e^{-j\beta_i}e^{-j2\pi\Delta f n}}}{\bigg{\}}},
 \label{equation:Function_CFO_1}
 \end{equation}
 in which the statistics $\hat \alpha_i$ and $\hat \beta_i$ are to be substituted. These are given by
 \begin{equation}
 \hat \beta_i = \angle {\bigg{\{}}\sum_{n=0}^{N-1}{r_i[n]s^\text{C}[n]e^{-j2\pi \Delta f n}}{\bigg{\}}}
 \label{equation:PO_Compact}
 \end{equation}
 \begin{equation}
 \hat \alpha_i = \frac{1}{N}{\bigg{|}}r_i[n]s^\text{C}[n]e^{-j2\pi\Delta f n}{\bigg{|}},
 \label{equation:Alpha_Compact}
 \end{equation}
 where the statistic $\hat \alpha_i$ follows from substituting \eqref{equation:PO_Compact} into \eqref{equation:Amplitude_Estimator}. Therefore, \eqref{equation:PO_Compact} and 
 \eqref{equation:Alpha_Compact} can be substituted into  \eqref{equation:Function_CFO_1} to yield the value of $\Delta f$ that maximizes the objective function. This can be written as
 \begin{equation}
 \hat{\Delta f} = \arg\,\underset{\Delta f}{\max}\,\left\{ \frac{1}{N}\sum_{i=1}^{\Lr}{{\left|\sum_{n=0}^{N-1}{r_i[n]s^\text{C}[n]e^{-j2\pi\Delta f n}}\right|^2}}\right\}.
 \end{equation}
 Accordingly, the ML estimator for the carrier frequency offset can be understood as the bin corresponding to the maximum of the sum of the squared-periodograms of the received signals, after a temporal matched filter. Since the periodogram is defined as the Fourier Transform of the autocorrelation, it is necessary to find its maximum numerically, by using the Fast - Fourier Transform (FFT). 
 
 The FFT yields a sampled version of the DTFT, such that there is no guarantee that the carrier frequency offset falls within a specfic integer bin of the FFT. Thus, the three largest points in the FFT can be found and parabolic interpolation can be performed thereafter. The motivation behind this approach is trying to find the real maximum of the function. The actual maximum is not generally found, but an accurate estimate can be found, instead.
 
 \subsection{Quadratic Interpolation of Spectral Peaks}
 
\begin{figure}%[!htb]
 	\begin{center}
 		\includegraphics[width = 12cm, height = 8cm]{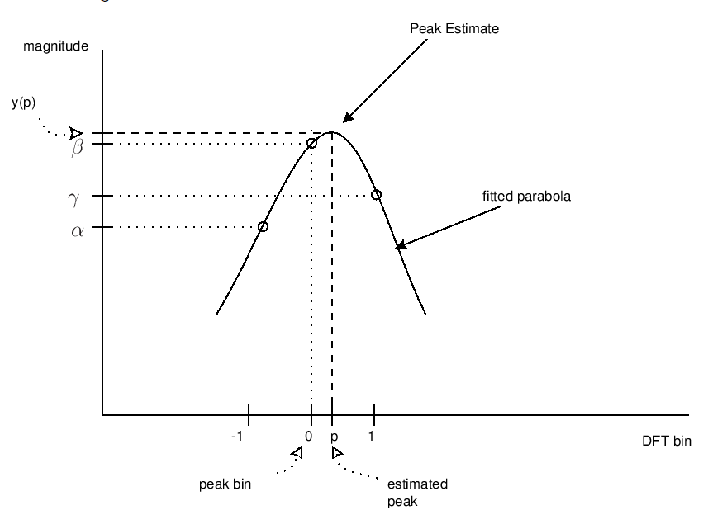}
 		\caption{Illustration of parabolic peak interpolation using the three samples closest to the peak.}
 		\label{fig:parab_interp}
 	\end{center}
 \end{figure}

 This subsection is meant to explaining how parabolic interpolation can be applied to obtain a better estimate of the carrier frequency offset.
 In general, assume that the shape of the Energy Spectral Density (ESD) of the received stochastic process is the one pictured in Figure~\ref{fig:parab_interp}.
 Then, the equation of a parabola is given by
 \begin{equation}
 y(x) \triangleq a(x-p)^2 + b,
 \end{equation}
where $p$ is the interpolated peak location (in FFT bins). At the three nearest samples, the images of the parabola at these points are
 \begin{equation}
 y(-1) = \alpha \qquad y(0) = \beta \qquad y(1) = \gamma.
 \end{equation}
 
 %%%%%%%%%%%%%%%%%%%%%%%%%%%%%%%%%%%%%%%%%%%
 
 %%%%%%%%%%%%%%%%%%%%%%%%%%%%%%%%%%%%%%%%%%%
 Then, expressing the three samples in terms of the interpolating parabola, it is found that
 \begin{equation}
 \begin{split}
 \alpha &= ap^2 + 2ap + a + b \\
 \beta &= ap^2 + b \\
 \gamma &= ap^2-2ap + a + b.
 \end{split}
 \end{equation}
 Therefore, the interpolated peak location in bins is given by
 \begin{equation}
 p = \frac{1}{2}\frac{\alpha - \gamma}{\alpha -2\beta + \gamma} \, \in {\big{[}}-\frac{1}{2},\frac{1}{2}{\big{]}},
 \label{FFTint}
 \end{equation}
 such that the amplitude at that bin is 
 \begin{equation}
 y(p) = \beta - \frac{1}{4}(\alpha - \gamma)p.
 \end{equation}
 Finally, by using \eqref{FFTint}, an estimate of the interpolated peak location can be obtained. If $k^*$ denotes the FFT bin that yields the maximum value, then $(k^*+p)$ yields the absolute position of the interpolated bin. Then, the analog frequency estimate is expressed as
 \begin{equation}
 \tilde{\omega} = \frac{k^*+p}{N}f_s,
 \end{equation}
where $f_s$ is the sampling frequency in Hertz and $N$ is the number of points in the FFT.
 
 \subsection{MLE for the Noise Variance}
 
 This subsection is devoted to find the MLE for the noise variance parameter, $\sigma^2$. Recalling the Log-Likelihood Function
 \begin{equation}
 \ln{p(\br; \bm \xi)} = -N\Lr\ln{\pi\sigma^2} - \frac{1}{\sigma^2}\|\br - \bM(\bm \alpha,\Delta f) \bt\|_2^2,
 \end{equation}
 the MLE for the noise variance can be easily found from
 \begin{equation}
 \frac{\partial \ln{p(\br; \bm \xi)}}{\partial \sigma^2} = -\frac{N\Lr}{\sigma^2} + \frac{1}{\sigma^4}\|\br - \bM(\bm \alpha,\Delta f) \bt\|_2^2.
 \end{equation}
 The condition for the derivative to vanish is that
 \begin{equation}
 \widehat{\sigma^2}  = \frac{1}{N\Lr}\|\br - \hat{\bM}(\bm \alpha,\Delta f) \bt\|_2^2,
 \end{equation}
 where $\widehat{\bM}(\bm \alpha,\Delta f)$ is the estimate of the matrix $\bM(\bm \alpha,\Delta f)$ that models how the signal energy is transferred from the transmitter to the receiver. The estimate $\widehat{\bM}(\bm \alpha,\Delta f)$ is computed using the ML estimators of the previous parameters.
 
 \subsection{MLE for the Signal to Noise Ratio}
 
 This last subsection is devoted to provide the ML estimator for the $\SNR$ metric. To find this statistic, the following property of ML estimation can be applied \cite{Kay:Fundamentals-of-Statistical-Signal:93}. 
 \begin{lemma}{\textbf{Invariance Property of the MLE}}:
 	The MLE of the parameter $\alpha = g(\xi)$, where the pdf $p(\bm x; \xi)$ is parameterized by $\xi$, is given by
 	\begin{equation}
 	\hat \alpha = g(\hat \xi),
 	\end{equation}
 	where $\hat \xi$ is the MLE of $\xi$.
 \end{lemma}
 This nice property allows us to find the MLE for $\SNR = \frac{\alpha_i^2}{\sigma^2}$, which is simply
 \begin{equation}
 \hat {\gamma_i} = \frac{\hat \alpha_i^2}{\widehat {\sigma^2}},
 \end{equation}
 where $\hat \alpha_i$ and $\widehat{\sigma^2}$ are the MLEs for $\alpha_i$ and $\sigma^2$ provided in the previous subsections.
	As a summary, the ML estimators for the parameters are
\begin{equation}
\begin{split}
\widehat {\Delta f}_\text{ML} &= \arg\,\underset{\Delta f}{\max\,}\quad \sum_{i=1}^{\Lr}{{\bigg{|}}\sum_{n=0}^{N-1}{s^\text{C}[n]r_{i}[n]e^{-j2\pi\Delta f n}}{\bigg{|}}^2} \\
\hat {\beta}_{i,\text{ML}} &= \angle\{\sum_{n=0}^{N-1}{s^\text{C}[n]r_{i}[n] e^{-j2\pi \Delta f n}}\} \\
\hat {\alpha}_{i,\text{ML}} &= \frac{1}{N}|\sum_{n=0}^{N-1}{s^\text{C}[n]r_{i}[n] e^{-j2\pi \Delta f n}}| \\
\widehat {\sigma^2}_\text{ML} &= \frac{1}{\Lr N}||\br - \hat {\bM}(\bm \alpha,\Delta f) \bt||_2^2 \\
\hat \gamma_{i,\text{ML}} &= \frac{\hat \alpha_i^2}{\widehat{\sigma^2}} \\
\widehat {\SNR}_\text{ML} &= \frac{\trace\{\widehat {\bP}\}}{\Lr\widehat {\sigma^2}},
\end{split}
\end{equation}

To assess the performance of the different estimators, some results are presented hereafter. In the simulations, a range from $-15$ dB to $10$ dB has been considered for $\SNR$, which is a typical range in millimeter wave communications. The amplitude parameters are generated from a ${\cal U}[0,1]$ distribution. The phase-offsets are generated from a ${\cal U}(0,2\pi)$ and the carrier frequency offset is generated as ${\cal U}[-1/2,1/2]$. The number of antennas is set to $\Nt = 32$ and $\Nr = 32$, and the number of RF chains employed at both transmitter and receiver are set to $\Lt = 4$ and $\Lr = 4$. The training sequence $t[n]$, $0\leq n \leq N-1$ contains normalized QPSK symbols, with $N = 64$ time-domain samples. The results have been averaged over $N_\text{MC} = 1000$ Monte Carlo realizations. 

\begin{figure}[htbp]	
	\centering
	{\includegraphics[width=14cm,height=8.5cm]{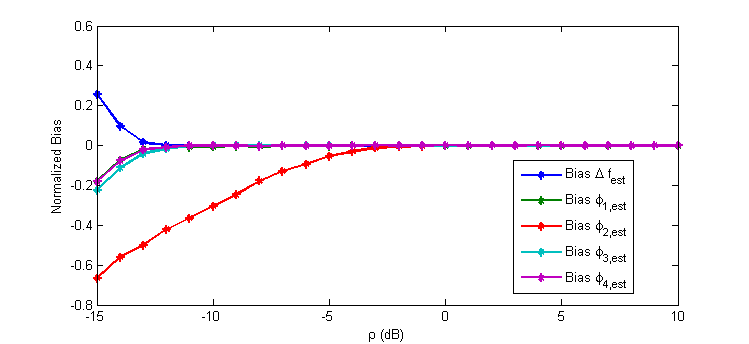}}
		\caption{Normalized sample bias for the angular parameters $\beta_i$, $i = 1,\ldots,4$ and $\Delta f$.}
 \label{fig:Nbias}
\end{figure}

The normalized sample bias of the the different parameters is shown in Figure~\ref{fig:Nbias}. As predicted by the Estimation Theory, at low $\SNR$ regime the sample bias of the parameters is non zero whereas it decreases to zero asymptotically with $\SNR$. The estimators are, thereby, asymptotically unbiased.

\begin{figure}[htbp]
	\centering
	{\includegraphics[width=14cm]{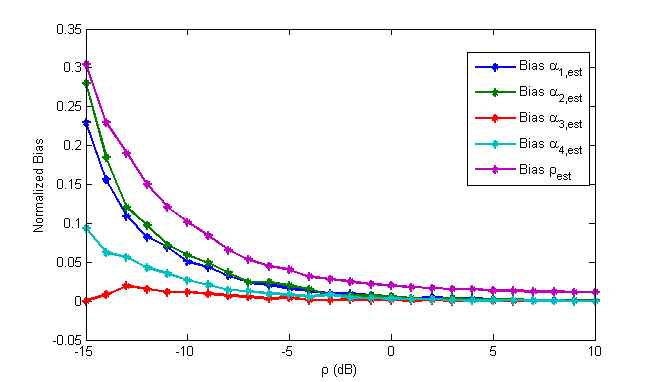}}
	\caption{Normalized sample bias for the amplitude and $\SNR$ parameters.}
\end{figure}

\begin{figure}[htbp]
		\centering
	{\includegraphics[width=14cm]{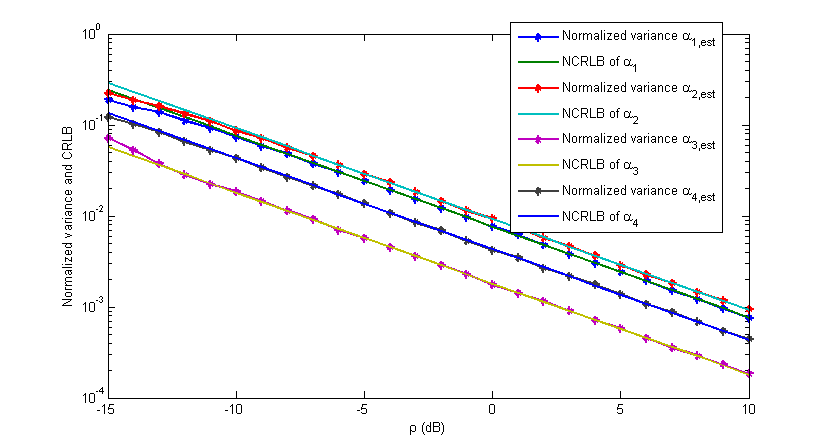}}
	\caption{Normalized sample variance and NCRLB for the amplitude parameters.}
\label{fig:Nvar_NCRLB_Alpha}
\end{figure}

\begin{figure}[htbp]
	\centering 
	{\includegraphics[width=14cm]{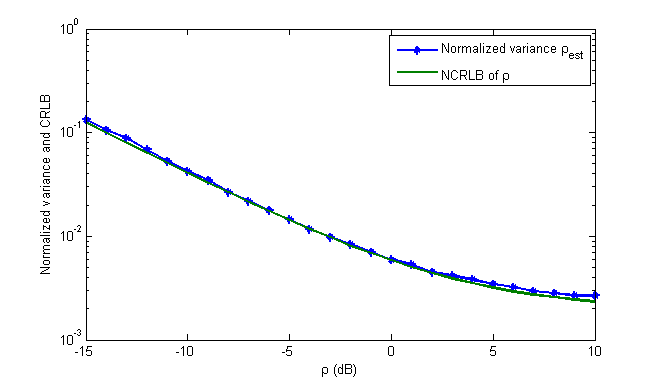}}
		\caption{Normalized sample variance and NCRLB for $\SNR$.}
	\label{fig:Nvar_NCRLB_SNR}
\end{figure}

\begin{figure}[htbp]
	\centering
	{\includegraphics[width=14cm]{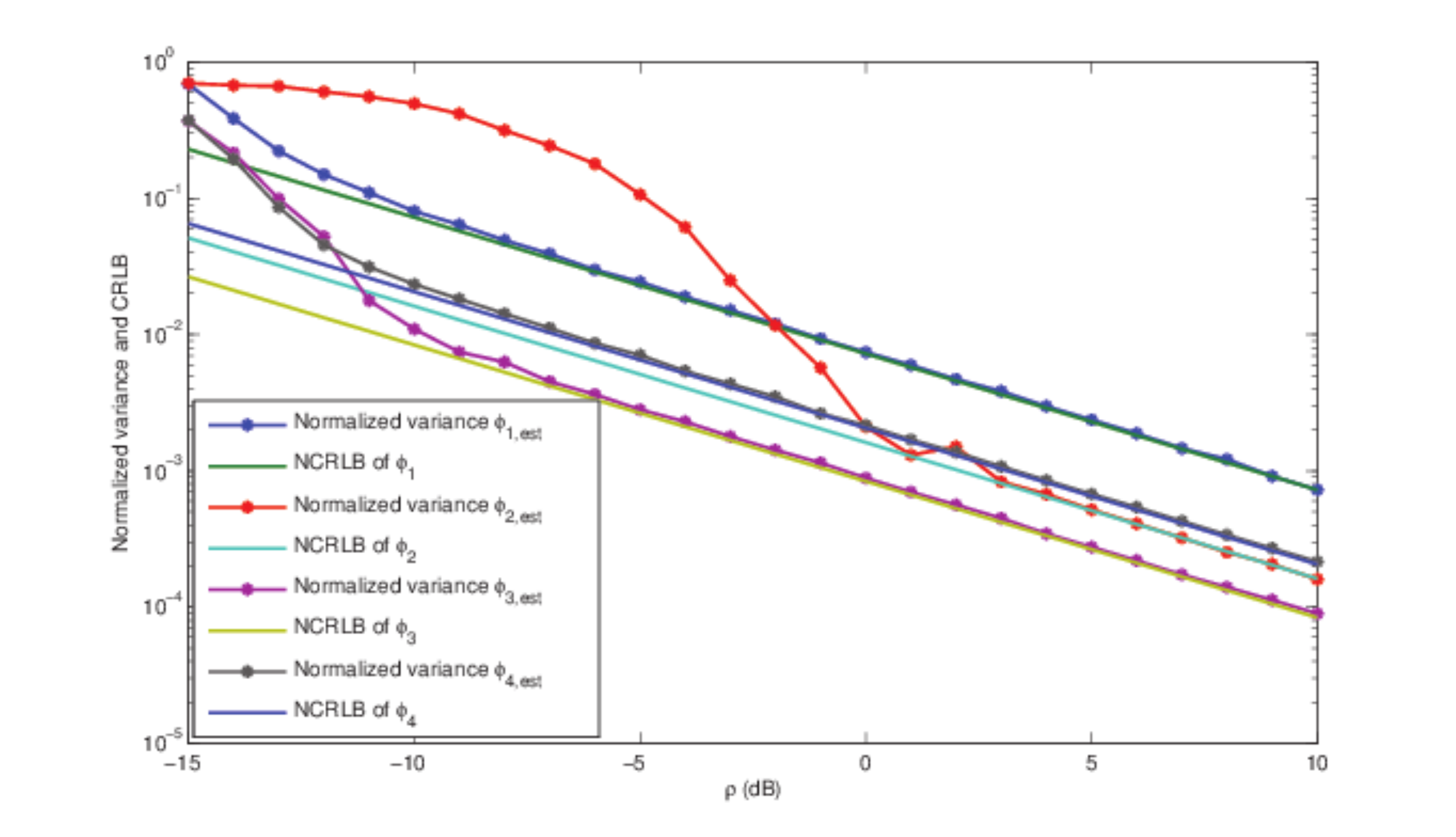}}	
	\caption{Normalized sampled variance and NCRLB for the phase offsets $\beta_i$, $i = 1,\ldots,4$.}
	\label{fig:Nvar_NCRLB_PO}
\end{figure}

\begin{figure}[htbp]
	\centering
	{\includegraphics[width=12cm]{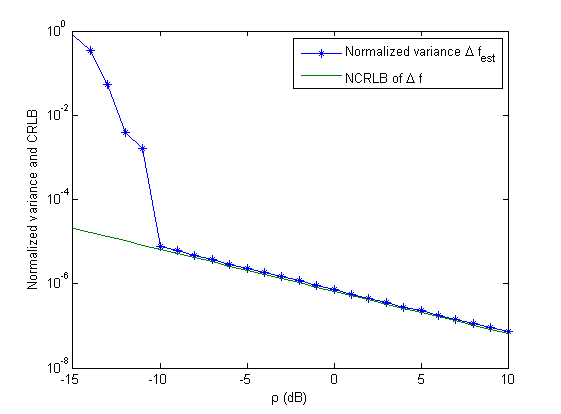}}
	\caption{Normalized sample variance and NCRLB obtained for the carrier frequency offset parameter.} \label{fig:Nvar_NCRLB_Deltaf}
\end{figure}

Now, the efficiency of the estimators is evaluated through the normalized sample variance. The normalized sample variances for the different estimators is shown in Figure~\ref{fig:Nvar_NCRLB_Alpha}, Figure~\ref{fig:Nvar_NCRLB_SNR}, Figure~\ref{fig:Nvar_NCRLB_PO} and Figure~\ref{fig:Nvar_NCRLB_Deltaf}, along with their corresponding normalized CRLBs. It can be noticed that the normalized sample variance of the estimators does not lie within the NCRLB for low values of $\SNR$, as it is expected. Nevertheless, for high values of $\SNR$, the estimators present a normalized sample variance that lies within the NCRLB. Therefore, the estimators are asymptotically efficient. 

%%%%%%%%%%%%%%%%%%%%%%%%%%%%%%%%%%%%%%%%%

\newpage
\bibliographystyle{IEEEtran}

\end{document}